\documentclass[trackchanges,twocolumn]{aastex7}

\usepackage{ulem}
\usepackage{amsmath}
\usepackage{graphicx}
\usepackage{natbib}
\usepackage{longtable}
\usepackage{supertabular,booktabs}
\usepackage{sidecap}
\usepackage{enumitem}
\usepackage{txfonts}
\usepackage{epstopdf}
\usepackage{tabularx}
\usepackage{ltablex}

\usepackage[normalem]{ulem}
\newcommand{\Msun}{M$_{\odot}$}

\begin{document}

\title{Detection of actinides in CEMP-rs stars}

\author[orcid=0009-0000-1753-2212,gname='Akkara Muhammed',sname='Riyas']{A. M. Riyas}
\affiliation{Department of Physics, University of Calicut, Thenhipalam, Malappuram, Kerala-673635, India}
\email{riyasbinzaid@gmail.com}  

\author[orcid=0000-0002-3532-2793,gname=Drisya, sname='Karinkuzhi']{D. Karinkuzhi} 
\affiliation{Department of Physics, University of Calicut, Thenhipalam, Malappuram, Kerala-673635, India}
\affiliation{Institut d'Astronomie et d'Astrophysique, Universit\'e Libre de Bruxelles, ULB, Campus Plaine C.P. 226, Boulevard du Triomphe, 1050 Bruxelles, Belgium}
\email{drdrisyak@uoc.ac.in}

\author[orcid=0000-0003-0499-8608,gname=Sophie,sname= Van Eck]{S. Van Eck}
\affiliation{Institut d'Astronomie et d'Astrophysique, Universit\'e Libre de Bruxelles, ULB, Campus Plaine C.P. 226, Boulevard du Triomphe, 1050 Bruxelles, Belgium}
\affiliation{BLU-ULB, Brussels Laboratory of the Universe, blu.ulb.be}
\email{sophie.van.eck@ulb.be}

\author[orcid=0000-0001-6159-8470,gname=Arthur,sname= Choplin]{A. Choplin}
\affiliation{Institut d'Astronomie et d'Astrophysique, Universit\'e Libre de Bruxelles, ULB, Campus Plaine C.P. 226, Boulevard du Triomphe, 1050 Bruxelles, Belgium}
\affiliation{BLU-ULB, Brussels Laboratory of the Universe, blu.ulb.be}
\email{Arthur.Choplin@ulb.be}

\author[orcid=0000-0002-9110-941X,gname=Stephane,sname= Goriely]{S. Goriely}
\affiliation{Institut d'Astronomie et d'Astrophysique, Universit\'e Libre de Bruxelles, ULB, Campus Plaine C.P. 226, Boulevard du Triomphe, 1050 Bruxelles, Belgium}
\affiliation{BLU-ULB, Brussels Laboratory of the Universe, blu.ulb.be}
\email{stephane.goriely@ulb.be}

\author[orcid=0000-0001-6008-1103,gname=Lionel,sname=Siess]{L. Siess}
\affiliation{Institut d'Astronomie et d'Astrophysique, Universit\'e Libre de Bruxelles, ULB, Campus Plaine C.P. 226, Boulevard du Triomphe, 1050 Bruxelles, Belgium}
\affiliation{BLU-ULB, Brussels Laboratory of the Universe, blu.ulb.be}
\email{lionel.siess@ulb.be}

\author[orcid=0009-0002-4119-983X,sname= Farha]{ A. A. Farha}
\affiliation{Department of Physics, University of Calicut, Thenhipalam, Malappuram, Kerala-673635, India}
\email{farhagaffur1616@gmail.com}


\begin{abstract}

The carbon-enhanced metal-poor stars with hybrid enrichments of slow- and rapid neutron-capture elements, the so-called CEMP-rs stars, still raise many questions due to their elusive abundance signatures. In our recent analysis, we found that heavy r-process elements are enhanced in these objects and can be explained by the intermediate neutron-capture process (i-process) occurring in low-mass, very low-metallicity asymptotic giant branch (AGB) stars. However, the origin of actinides such as thorium and uranium is typically associated with explosive nucleosynthesis in highly neutron-rich environments, and their detection in stellar spectra remains challenging due to severe line blending from 
other elements and carbon-bearing molecules.
In this work, we investigate the presence of thorium and uranium abundances in a sample of  CEMP-rs stars using their high-resolution spectra obtained with the UVES spectrograph mounted on the UT2 (Kueyen) ESO VLT. Thorium is robustly detected in three stars, while uranium remains marginally detected, allowing only upper limits to be derived. Comparison with theoretical i-process nucleosynthesis models demonstrates that the observed abundances can be reproduced within uncertainties, supporting an i-process origin for these elements. This study reports the first detection of actinides in stars confirmed as CEMP-rs stars, providing new constraints on their nucleosynthetic history. Furthermore, these detections provide a potential way to estimate in the future the time elapsed since the proton-ingestion episode in AGB stars using cosmochronometry techniques, and more generally to place lower limits on the ages of the resulting white dwarf remnants.
\end{abstract}


\section{Introduction}
\label{Sect:Intro}

The low- to intermediate-mass asymptotic giant branch (AGB) stars at different metallicities play a crucial role in producing half of the elements heavier than iron by the slow neutron-capture process (s-process). 
Mass-transfer episodes from these stars are well established as the origin of the s-process enrichment observed in many extrinsically-enriched stars, such as CH, barium, and Carbon-Enhanced Metal-Poor (CEMP)-s stars. However, the origin of the peculiar subgroup of CEMP stars exhibiting hybrid enhancements of elements typically produced by two distinct neutron-capture processes, the slow neutron-capture process (s-process) and the rapid neutron-capture process (r-process) 
\citep{Jonsell2006,Masseron2010,Lugaro2023}, remains a long-standing problem. These objects, known as CEMP-rs stars, still lack a clear explanation for their formation.
Recent studies \citep{Hampel2016,Hampel2019,Karinkuzhi2021,Choplin2021,Choplin2024} identified that the proton ingestion episodes (PIEs) in AGB stars, which activate the intermediate neutron-capture process (i-process) at conditions intermediate between those required by the s- and r-processes, can be held responsible for the hybrid enrichment in CEMP-rs stars.
Even the presence of hybrid enrichment in stars with moderately higher metallicities ($[\mathrm{Fe/H}] \approx -$0.7 to $-$0.1)
 could also be explained by the i-process occurring in AGB stars under certain modified conditions of mixing in their interiors \citep{Herwig2011,Karinkuzhi2023,Choplin2024}. 
 However, the presence of heavy r-process elements in CEMP-rs stars has remained largely unexplored, mainly owing to the lack of spectra at bluer wavelengths.
  Recently, \citet{Riyas2026} measured the abundances of heavy r-process elements in these stars,  
such as terbium, holmium, thulium, ytterbium, lutetium, tantalum, and iridium.
Usually, the r-process is considered 
as the main
mechanism responsible for the 
origin of these elements
in the Galaxy. However, our analyses indicate that their overproduction in CEMP-rs stars 
can be explained by the i-process occurring in low-mass, low-metallicity  AGB stars. 

Similar situations exist for the actinides, such as thorium (Th) and uranium (U).  Producing actinides requires high neutron densities and has therefore long been thought to occur exclusively during r-process nucleosynthesis. Their detection in stars is therefore considered as a powerful probe of the physical conditions under which they are synthesized \citep{Hill2002,Frebel2007,Roederer2009}. 
In addition, Th and U abundances can allow us to estimate the ages of stars through cosmochronometry, providing insights into the timeline of early Galactic chemical evolution \citep{Cowan1999,Cayrel2001,Schatz2002}.
However, their measurement is very challenging, especially in the case of carbon stars, due to their weak absorption lines and blending from carbon-bearing molecules. Until now, most successful Th and U measurements have been made in very metal-poor r-process enhanced stars such as carbon normal r-II stars \citep{Cowan2002,Hill2002,Yushchenko2005,Roederer2009,Ren2012,Siqueira2014,Qianfan2024}. However, \citet{Gull2018} tentatively detected a Th II line
 in the very metal-poor star RAVE J094921.8-161722 
 which exhibits a hybrid s- and r- enrichment.
From this line, they derived a thorium abundance and subsequently attributed the measured overabundances to two independent s- and r-process sites, rather than to a single nucleosynthetic event.

In the present work, we investigate a sample of CEMP stars that were previously confirmed as CEMP-rs stars,  using the most recent non-local thermodynamic version of the TURBOSPECTRUM spectral synthesis code \citep{Gerber2023} to detect the presence of actinides in these stars. Using high-resolution and high signal-to-noise spectra, we carefully analyze the key spectral lines of Th II and U II, estimate their abundances where possible, and discuss the uncertainties involved. 

The structure of the paper is as follows: In Section ~\ref{Sample selection}, we describe the selection of targets and the data reduction procedure. Section ~\ref{Sect:parameters} outlines the methods used for the abundance analysis, including line selection and fitting techniques. In Section ~\ref{Sect:lines}, we present the results of thorium and uranium 
line measurements. In Section ~\ref{Sect:nucleosynthesis}, we discuss the implications of our findings for the nucleosynthesis processes in CEMP-rs stars and compare them with theoretical models. In Section ~\ref{Sect:ages}, we present the difficulties of using Th and U abundance as chronometers in CEMP-rs stars. Finally, we summarize our conclusions in Section ~\ref{Sect: conclusion}.

\begin{table}
\centering
\caption{Stellar parameters of the program stars}
\label{Tab:program_stars}
\begin{tabular}{lcccc}
\hline
\hline
 & $T_{\rm eff}$ & $\log g$ & $\xi$ & [Fe/H] \\
 & (K) & (cm~s$^{-2}$) & (km~s$^{-1}$) &  \\
\hline

\multicolumn{5}{c}{\textbf{HD 187861}} \\
LTE (a)  & $5000^{\pm 100}$ & $1.50^{\pm 0.25}$ & $2.00^{\pm 0.20}$ & $-2.60^{\pm 0.10}$ \\
NLTE (b) & $5200^{\pm 50}$  & $2.00^{\pm 0.20}$ & $2.00^{\pm 0.20}$ & $-2.40^{\pm 0.15}$ \\

\multicolumn{5}{c}{\textbf{HD 224959}} \\
LTE (a)  & $4969^{\pm 64}$  & $1.26^{\pm 0.29}$ & $1.63^{\pm 0.14}$ & $-2.36^{\pm 0.09}$ \\
NLTE (b) & $5100^{\pm 100}$ & $2.00^{\pm 0.10}$ & $1.63^{\pm 0.14}$ & $-2.26^{\pm 0.12}$ \\

\multicolumn{5}{c}{\textbf{CS 22891$-$171}} \\
LTE (a)  & $5215^{\pm 68}$  & $1.24^{\pm 0.09}$ & $2.14^{\pm 0.14}$ & $-2.50^{\pm 0.10}$ \\
NLTE (b) & $5300^{\pm 85}$  & $1.80^{\pm 0.20}$ & $2.14^{\pm 0.14}$ & $-2.19^{\pm 0.14}$ \\

\hline
\end{tabular}
\tablecomments{(a) \cite{Karinkuzhi2021}; (b) This work.}
\end{table}


\section{ Selection of targets and the data analysis}
\label{Sample selection}
In this study, we focus on  CEMP-rs stars initially analysed by  ~\citet{Karinkuzhi2021}. We use 
near-ultraviolet spectra taken with the UVES instrument on the Very Large Telescope (VLT) at the European Southern Observatory (ESO). These data come from our observing programs (105.20LJ.001 and 105.20LJ.002) and from the ESO archives. The spectra have a resolution of 
$R \sim 47 000$ and cover wavelengths from around 
3280~\AA\ to 6835~\AA, allowing us to study a wide range of elements in detail. The 
abundance
profiles
of these stars, 
derived from an LTE analysis, are presented in 
\cite{Riyas2026}.


In the present study the stellar parameters (Table~\ref{Tab:program_stars}) were determined 
using NLTE-corrected abundances following \citet{Gerber2023}, using publicly available model atoms and departure-coefficient grids from the NLTE MPIA tools\footnote{\url{https://nlte.mpia.de}}. 
 NLTE model atoms come from the following studies: 
 H \citep{Mashonkina2008}, 
 O \citep{Bergemann2021}, 
 Na \citep{Larsen2022}, 
 Mg \citep{Bergemann2017}, 
 Ca \citep{Mashonkina2017, Semenova2020},
 Ti \citep{Bergemann2011},
 Mn \citep{Bergemann2019}, 
 Fe \citep{Bergemann2012, Semenova2020}, 
 Co \citep{Bergemann2010, Yakovleva2020}, 
 Ni \citep{Bergemann2021, Voronov2022}, 
 Sr \citep{Bergemann2012a}, 
 Ba \citep{Gallagher2020}
 Y \citep{Storm2023}, 
 Eu \citep{Storm2024}.

\begin{figure}
\centering
\includegraphics[width=0.45\textwidth]{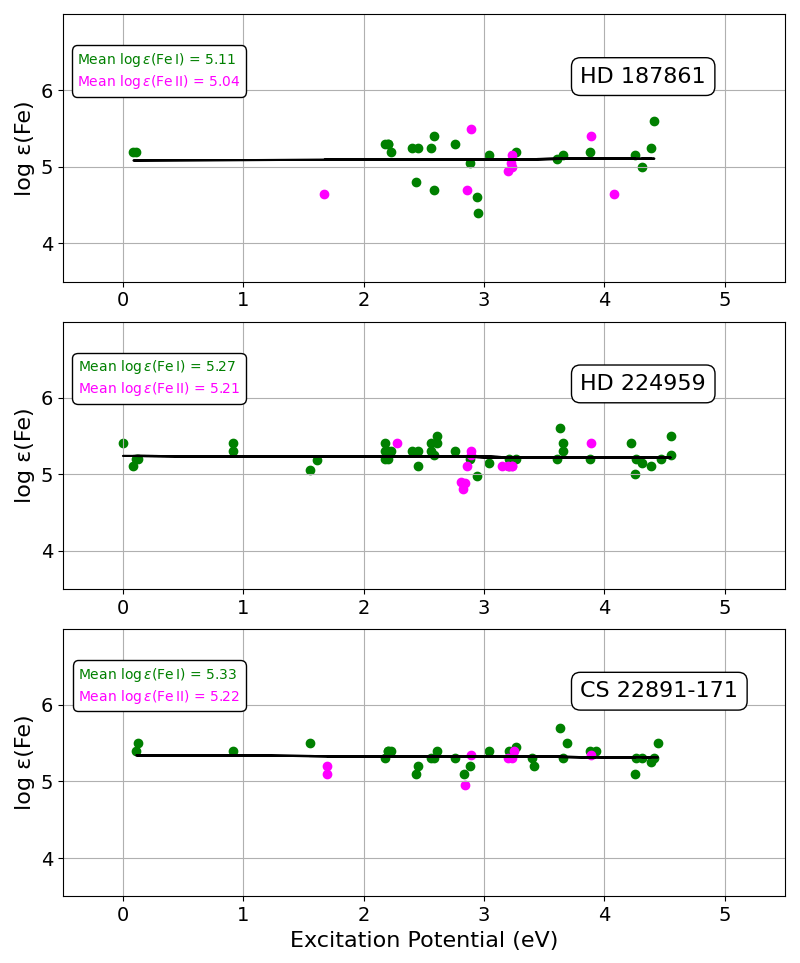}
\caption{Abundance vs excitation potential of Fe lines for the stars HD 187861, HD 224959, and CS 22891-171 (from top to bottom). Green and magenta circles represent Fe I and Fe II abundances, respectively. The straight line indicates the linear fit to the data.}
\label{Fig:Exitation}
\end{figure}

The stellar parameters derived under the LTE 
assumption were adopted as initial values and subsequently iterated after determining the Fe I and Fe II abundances at NLTE conditions.
Only weak lines, defined as those with a line-to-continuum flux ratio below 0.2, were considered for the Fe abundance determination, 
to minimise the impact of line saturation, microturbulence, and uncertainties in line broadening. The excitation and ionization equilibria were further examined to refine the T$_{\rm eff}$ and log $g$ values. The process was repeated until excitation and ionization balance were achieved, as indicated by a near-zero slope in the abundance trends 
(Fig.~\ref{Fig:Exitation}). The metallicity 
was derived based on the NLTE Fe abundances. 

Multiple weak \ion{Fe}{1} lines with low equivalent widths were selected to avoid the influence of line saturation. 
As the number of weak Fe lines with NLTE corrections are limited, we adopted the microturbulent velocity ($\xi$) derived using the LTE Fe abundances, by enforcing no correlation between the derived abundances and line strengths (i.e., reduced equivalent widths), thereby ensuring a consistent abundances across the selected lines.  Furthermore, we do not expect the $\xi$ values to change significantly under NLTE conditions, as the Fe I abundances, especially the weak lines, remain largely unaffected across the parameter range relevant to our objects \citep{Lind2012}.   We also measured the sensitivity of the derived abundances to microturbulence, 
by varying the microturbulent velocity by $\pm$0.5 km s$^{-1}$. We note that this variation did not produce significant changes in the derived abundances.

 \section{Abundance analysis}
\label{Sect:parameters}
The elemental abundances derived for the three CEMP-rs stars are presented in Table~\ref{Tab:abundances}. Abundances were determined from the best available spectral lines, which in some cases include only a single measurable transition. For a few elements, the usable lines lie toward the blue end of the spectrum where the signal-to-noise ratio 
is lower ($\mathrm{S/N} < 20$ at $\lambda \approx 3500\,\text{\AA}$) and blending effects are more significant. In cases where the features were too weak, blended, or affected by noise, only upper limits or uncertain abundances could be determined; these cases are clearly indicated in Table ~\ref{Tab:abundances}.
For the elements O, Na, Mg, Ca, Ti, Mn, Co, Ni, Sr, Y, Ba, and Eu, non-LTE corrections were applied to the lines listed in Table~\ref{Tab:Linelists}.
The uncertainties in elemental abundances are derived as described in  Appendix C.

\section{Thorium and Uranium 
}
\label{Sect:lines}
\begin{deluxetable}{ccc}
\tabletypesize{\small}
\tablecaption{Derived thorium abundances from individual lines, and mean Th abundances, for the program stars.\label{Tab:Th_abundances}}
\tablehead{
\colhead{$\lambda$ (\AA)} &
\colhead{$\log \varepsilon_{\mathrm{Th}}$} &
\colhead{
$\langle \log \varepsilon(\mathrm{Th}) \rangle$}
}
\startdata
\multicolumn{3}{c}{\textbf{HD~187861}} \\
3469.921 & $-0.70$ & \\
4019.129 & $-0.85$ & \\
4086.521 & $-0.75$ & $-0.76$ \\
\multicolumn{3}{c}{\textbf{HD~224959}} \\
4019.129 & $-0.95$ & \\
4086.521 & $-0.85$ & $-0.90$ \\
\multicolumn{3}{c}{\textbf{CS~22891-171}} \\
3433.999 & $-0.55$ & \\
3435.977 & $-0.60$ & \\
3539.587 & $-0.70$ & \\
3469.921 & $-0.65$ & $-0.63$ \\
\enddata
\end{deluxetable}

\begin{figure}
\centering
\includegraphics[width=0.45\textwidth]{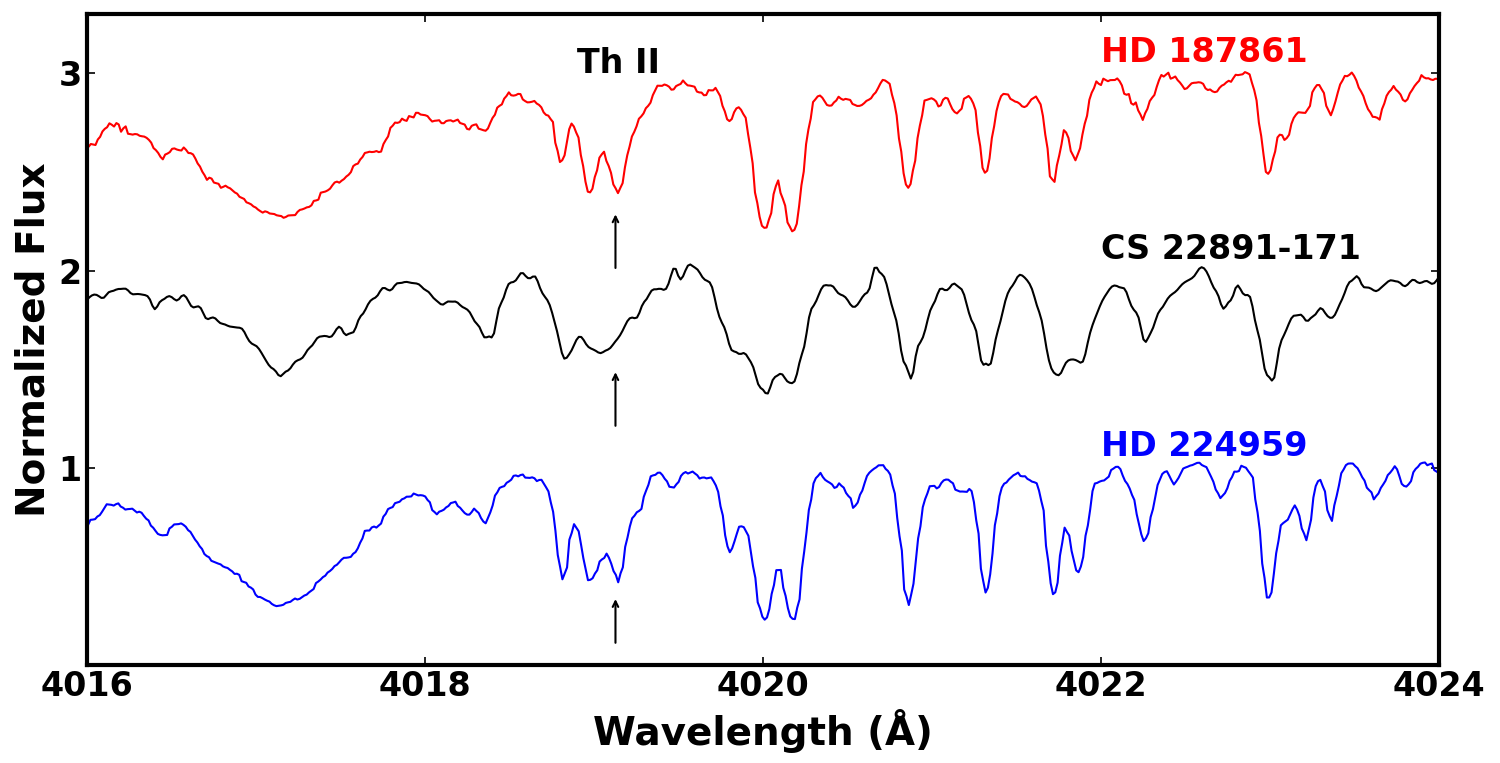}
\caption{Normalized spectra of the three CEMP-rs stars in the region around the Th II 4019.13~\AA~line. Fluxes are vertically shifted by one unit for clarity.} 
\label{Fig:spectra}
\end{figure}

\begin{figure}
\centering
\includegraphics[width=0.45\textwidth]{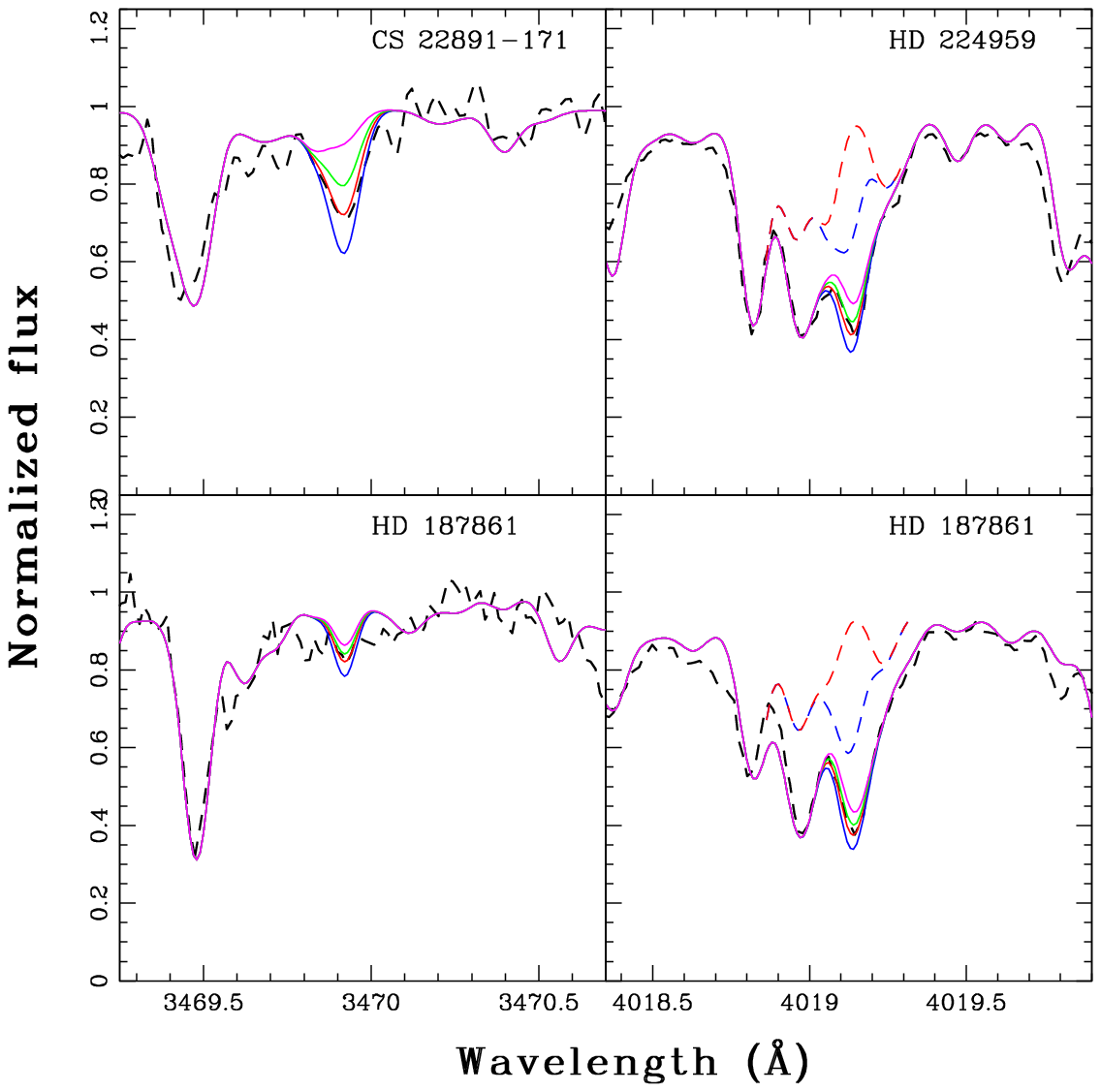}
\caption{Spectral fitting of the Th~II lines at 3469.92~\AA\ (left panel) and 4019.13~\AA\ (right panel) is shown for the CEMP-rs stars HD~187861, HD~224959, and CS~22891-171. Red lines correspond to spectral syntheses with the adopted Th~II abundances of $-$0.65 for CS~22891-171 and $-$0.70 for HD~187861  in the left panels, and $-$0.85 dex for HD~187861 and $-$0.95 dex  for HD~224959 in the right panels. Blue and green lines correspond to syntheses with abundances deviating by $\pm$ 0.3 dex from the adopted abundance. The black dashed line represents the observed spectrum. The magenta line corresponds to the synthesis with a null Th abundance. The blue dashed line represents the synthetic spectrum excluding $^{13}$CH contributions, while the red dashed line excludes both $^{13}$CH and Th~II contributions. }
\label{Fig:Thorium4019}
\end{figure}

\begin{figure}
\centering
\includegraphics[width=0.45\textwidth]{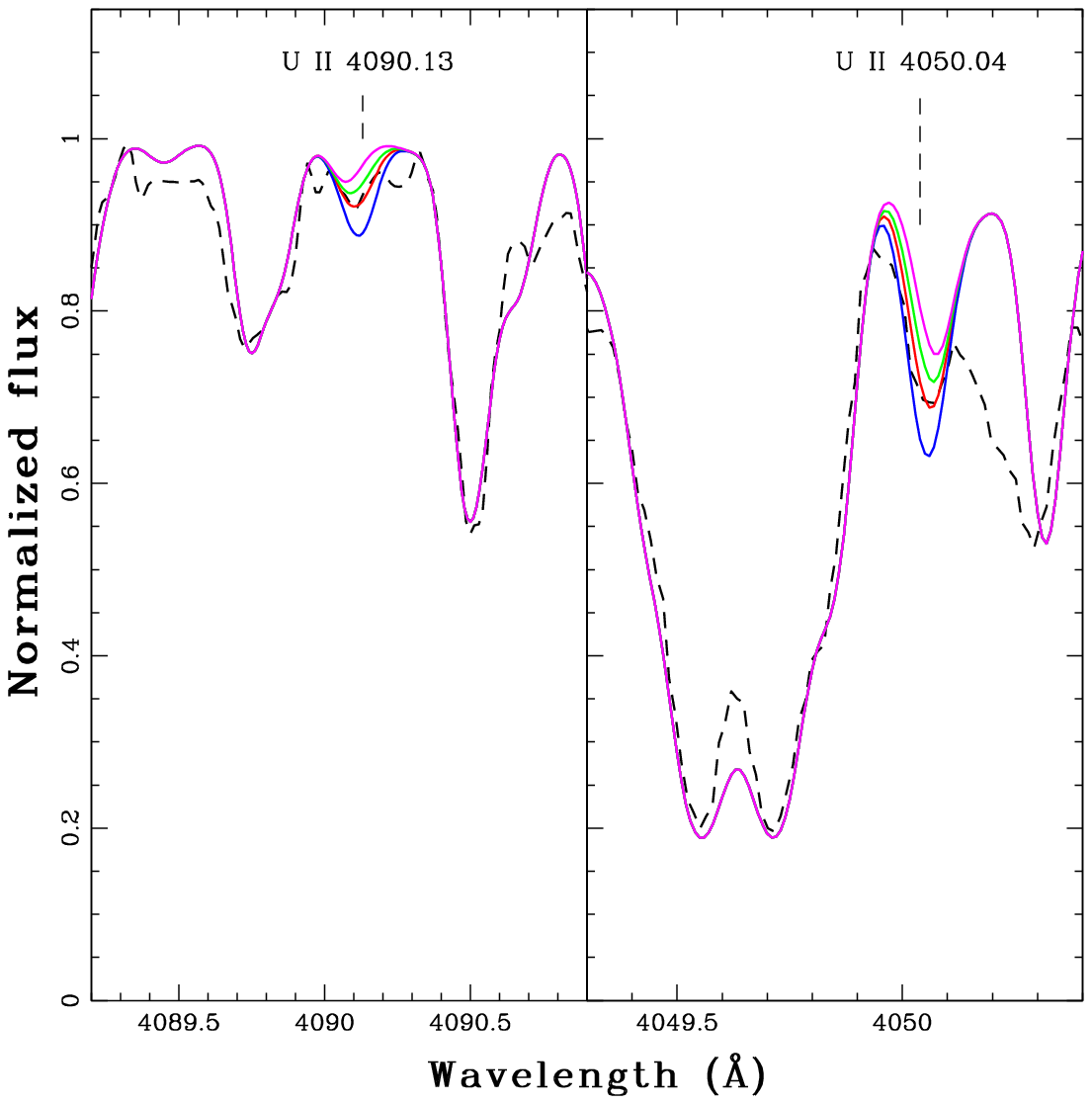}
\caption{The spectral fits for the U II lines at 4090.13~\AA\ and 4050.04~\AA\ are presented for CS~22891-171 in the left and for HD~187861 in the right panels, respectively. Red lines correspond to spectral syntheses with the adopted U~II abundances of $-$1.35 dex for CS~22891-171 and $-$1.40 dex for HD~187861. The blue, green, magenta, and
black curves have the same meaning as in Figure~\ref{Fig:Thorium4019}. }
\label{Fig:Uranium}
\end{figure}

Figure~\ref{Fig:spectra} shows the observed normalized spectra of the program stars in the region around the Th II 4019.13~\AA~line.
Figure~\ref{Fig:Thorium4019} presents the spectral synthesis fits of a few thorium lines in these CEMP-rs stars, while Figure~\ref{Fig:Uranium} shows the spectral synthesis fits of the uranium lines. The derived thorium abundances in our program stars corresponds to [Th/Fe] $\in[1.3;1.6]$. These values lie at the upper end, but are comparable, within uncertainties, to those measured in r-process enhanced stars 
(typically 
[Th/Fe] $\in [0.5; 1.8])$ \citep{Cowan2002,Hill2002,Frebel2007,Roederer2009}. This similarity suggests that the processes enriching these stars are capable of producing thorium at levels comparable to those seen in r-process–enhanced stars.
We have performed a 
synthesis of the spectral regions surrounding several \ion{Th}{2} lines in our CEMP-rs stars, resulting in measurable thorium abundances, as listed in Table~\ref{Tab:Th_abundances}. 
These results confirm the presence of thorium in all three stars.

In the present analysis, only upper limits could be placed on the uranium abundances, as the 
\ion{U}{2} lines at 3859.571~\AA, 4050.041~\AA\ and 4090.132~\AA, which are generally used for deriving U abundance in actinide boost stars \citep{Shah2023} could not be clearly detected in any of our program stars. These features are intrinsically weak and are further affected by blending with strong CN and \ion{Fe}{1} lines in the surrounding spectral regions, making accurate measurements challenging. 
Despite this, the derived upper limits on uranium still offer valuable constraints, as they could be used to estimate lower limits on the time since their production 
through radioactive chronometry, as discussed in Sect.~\ref{Sect:ages}. 

\section{Comparison with nucleosynthesis predictions}
\label{Sect:nucleosynthesis}
CEMP stars with both r- and s-process enhancements can be interpreted either as the result of a superposition of independent r- and s-process contributions, with pre-enrichment of the natal cloud followed by mass transfer from an AGB companion \citep{Jonsell2006,Bisterzo2012,Lugaro2012,Abate2016}, or as the outcome of a single nucleosynthetic event via the intermediate neutron-capture (i-process), which can reproduce the observed heavy-element patterns \citep{Cowan1977, Hampel2016,Denissenkov2017}. In the present work, we consider whether the i-process alone can account for the chemical peculiarities of our sample stars, and in particular their Th and U abundance constraints.
The heavy-element enrichment 
measured in CEMP-rs stars may indeed originate from a now-extinct AGB companion that transferred nucleosynthetically processed material through stellar winds \citep[e.g.][]{Lugaro2012, Bisterzo2012, Abate2013, Karinkuzhi2021, Choplin2024}. 
To test this scenario, we compare the derived abundances with predictions from AGB nucleosynthesis models computed using the \textsc{STAREVOL} code \citep{Siess2000, Siess2008}. 
Our AGB models follow the i-process during proton ingestion episodes (PIE) with a nuclear network comprising 1160 nuclei linked by 2123 reactions. 
We consider models with initial masses in the range $1\,M_\odot < M_{\rm ini} < 2\,M_{\odot}$ and metallicities $-3 \leq \mathrm{[Fe/H]} \leq -2$, consistent with those of the program stars. 
Further details on the modeling can be found in \citet{Choplin2021} and \citet{Choplin2022}.

\begin{figure}
\centering
\includegraphics[trim=2cm 0.5cm 1cm 0,width=0.50\textwidth]{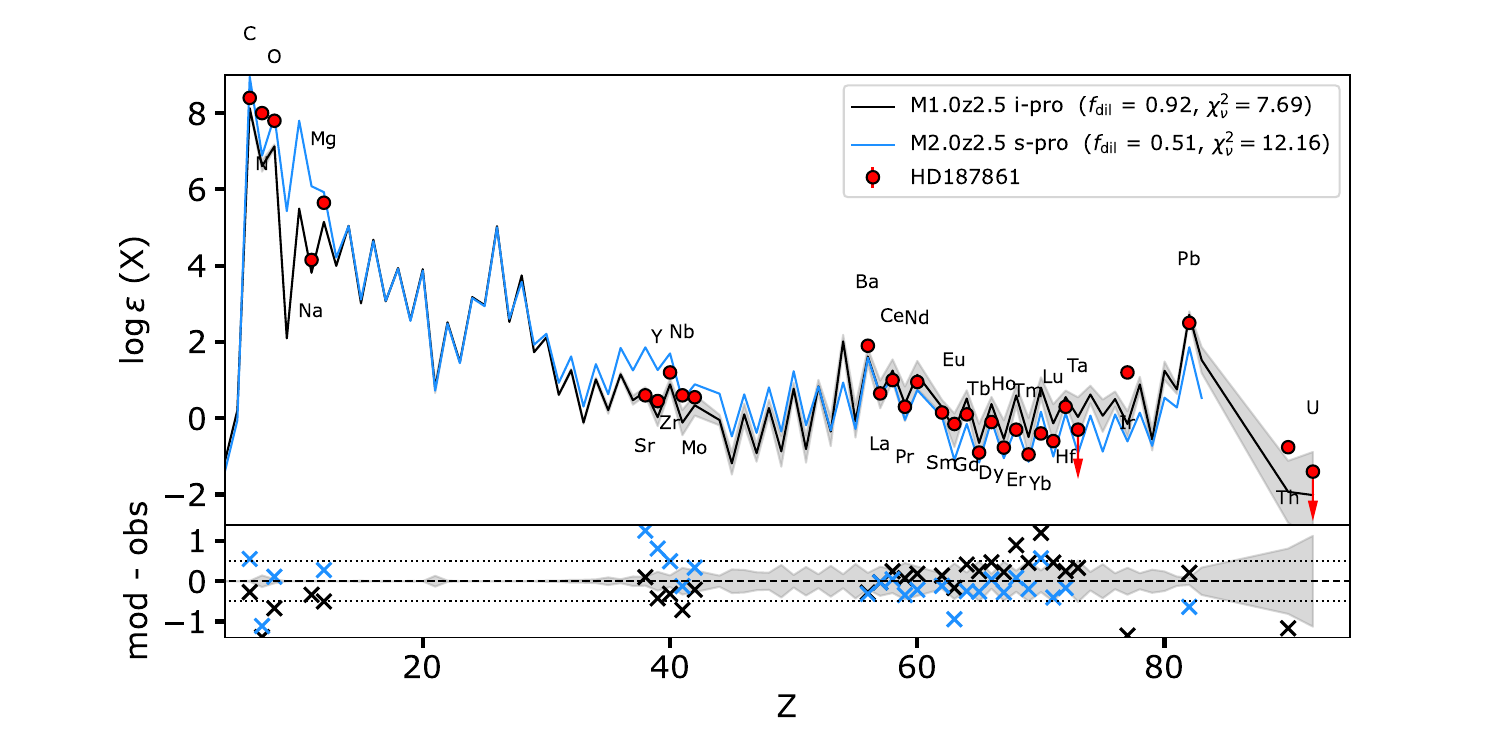}
\includegraphics[trim=2cm 0.5cm 1cm 0,width=0.50\textwidth]{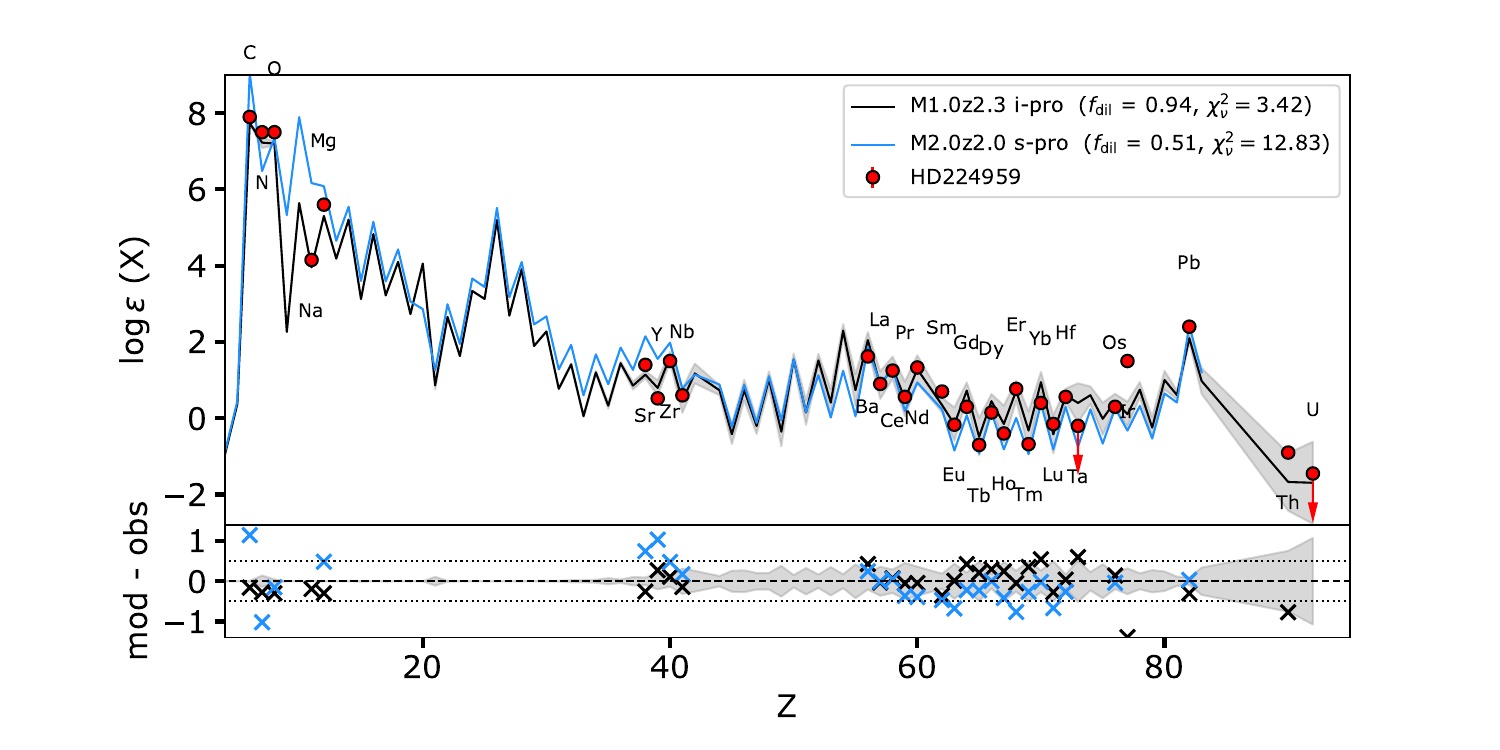}
\includegraphics[trim=2cm 0.5cm 1cm 0, width=0.50\textwidth]{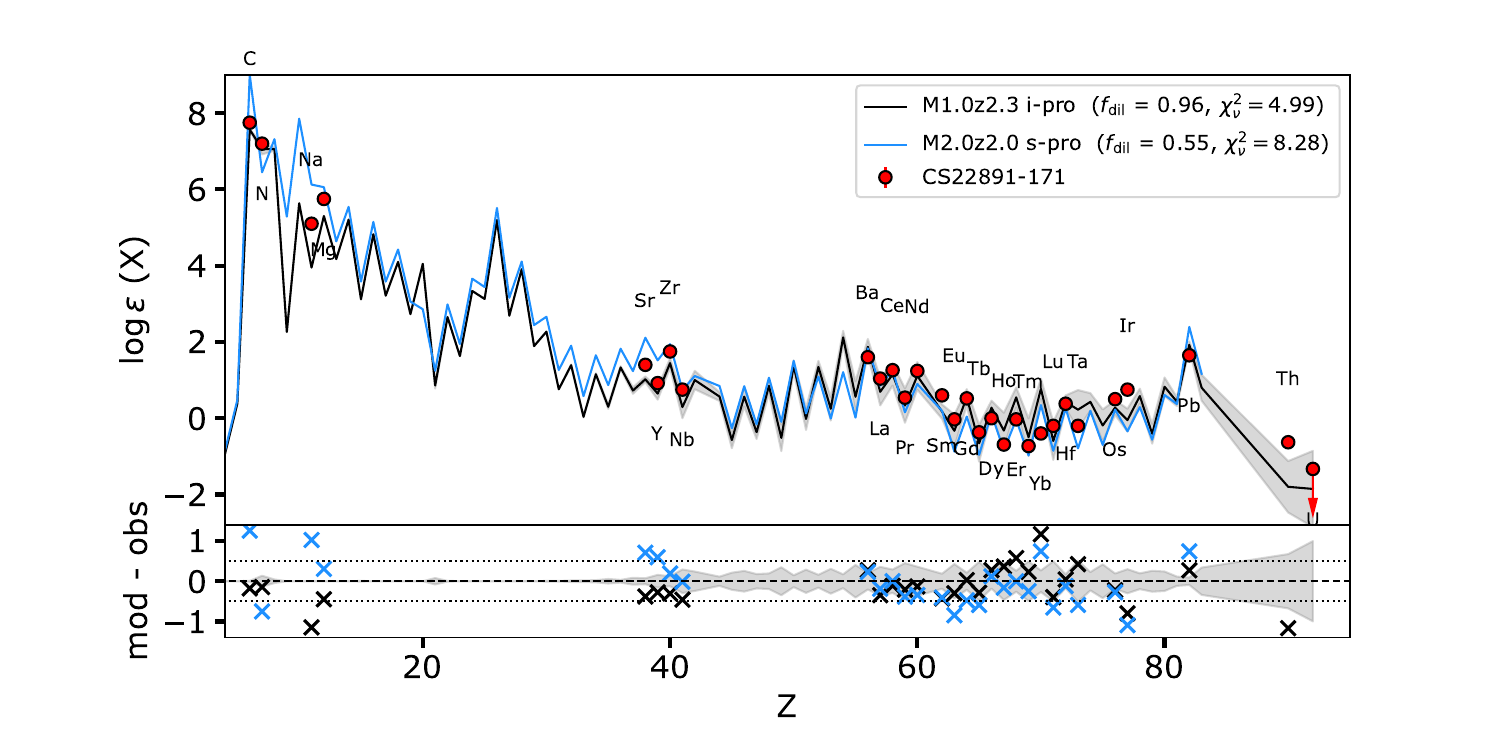}
\caption{The abundance patterns of the 3 CEMP-rs stars are compared with nucleosynthesis predictions from the STAREVOL code. The measured abundances are indicated by red circles.  
Upper limits are indicated with downward arrows. In all cases, the best-fitting theoretical predictions for both the i-process (black) and s-process (blue) are displayed. The grey shaded area corresponds to the typical i-process nuclear parameter uncertainties computed in \cite{Martinet2024} and based on a 1~\Msun, [Fe/H]~$=-2.5$ model.
}
\label{fig:pattern}
\end{figure}

To reproduce the chemical abundances of our sample, we follow the fitting procedure described in \citet[][Sect.~6.2]{Choplin2021}. 
For each star, the best-fitting AGB model is determined by minimizing the $\chi^2$ statistic, which quantifies the discrepancy between the measured and predicted abundances. 
The minimum $\chi^2$ is obtained by mixing a fraction of the AGB ejecta into the envelope of the companion star. 
The dilution factor, $f_{\rm dil}$, which is allowed to vary freely between 0.5 and 1, controls the amount of AGB material mixed into the envelope. 
We arbitrarily adopt a minimum dilution factor of 0.5 to avoid unrealistically small values, which would require the accretion of an implausibly large amount of AGB ejecta \citep{Choplin22cor}. 
The results of the fit are shown in Fig.~\ref{fig:pattern} for our three program stars. The best-fitting i-process abundance pattern is shown in black. 
For comparison, the best-fitting s-process pattern (blue), based on AGB models experiencing only s-process nucleosynthesis \citep{Goriely18c}, is also displayed.

The best-fitting i-process models correspond to 1\Msun\ AGB stars at $\mathrm{[Fe/H]} = -2.3$ and $-2.5$, whereas the best-fitting s-process models correspond to 2\Msun\ AGB stars at $\mathrm{[Fe/H]} = -2.0$ and $-2.5$. 
For all three program stars, the i-process models yield lower reduced $\chi^{2}$ values\footnote{The reduced $\chi^{2}$ is defined as $\chi_{\nu}^{2} = \chi^{2}/N_{\rm ab}$, where $N_{\rm ab}$ is the number of abundance measurements.} (see Fig.~\ref{fig:pattern}), in agreement with their classification as CEMP-rs stars. 
Notably, the s-process models systematically overproduce the Sr--Nb elements. 
The abundance pattern of the CEMP-rs star HD~224959 is particularly well reproduced by an i-process AGB model (with the exception of Ir), making this star a strong candidate for an i-process 
pollution.

While the i-process models reproduce most of the derived abundances within nuclear uncertainties (Fig.~\ref{fig:pattern}), there are notable discrepancies. Iridium ($Z=77$) is systematically underproduced by the i-process models by $0.5$--$1$~dex, but the spectroscopic abundance determination is uncertain for the three program stars (see Table~\ref{Tab:abundances}). In contrast, for HD~187861 and CS~22891$-$171, ytterbium ($Z=70$) is overproduced by the i-process models by approximately $1$~dex, 
but here again the abundance measurement is noted as uncertain in Table~\ref{Tab:abundances}. 
In addition, the upper limits derived for tantalum in HD~187861 and HD~224959 are only marginally compatible with our predictions and deserve further investigation, as Ta is generally produced in significant amounts during i-process nucleosynthesis \citep{Choplin2021}. 
For light elements, the residuals with the i-process models remain within $\sim 0.5$~dex  (with the exception of Na in CS~22891$-$171, see below), which is acceptable given that CEMP-rs stars may have experienced internal evolutionary processes (e.g. the first dredge-up) that can modify surface CNO abundances.
Na in CS~22891$-$171 is overestimated (underestimated) by s-process (i-process) models, suggesting that a unified i+s AGB model accounting for both nucleosynthesis mechanisms may be a solution, although such self-consistent models do not yet exist and current approaches treat the processes separately.

The lower limits derived for U are consistent with the model predictions, while the 
measured
Th abundances 
are only marginally reproduced by the best-fitting model, even when accounting for the nuclear uncertainties estimated by \citet{Martinet2024}. 
Beyond nuclear uncertainties, the predicted Th and U abundances are also sensitive to the spatial and temporal resolution of the stellar models, leading to variations of up to $\sim 1$~dex \citep{Choplin2022letter}. 
We note that higher Th abundances, closer to the 
measured values, can be obtained by adopting smaller dilution factors; however, this significantly degrades the fit to other elements. 
The key point is that i-process AGB models are capable of synthesizing Th and U without the need for an additional r-process operating in a different site.
Robust quantitative predictions for these elements will require substantial progress, particularly in the underlying nuclear physics. Finally, neutron-induced and/or spontaneous fission are not included in the present calculations and may further affect these results by transforming part of the actinides into lighter elements.



\section{Chronometers and Ages}
\label{Sect:ages}

If the thorium detected at the surface of CEMP-rs stars is supplied by an AGB companion that experienced i-process nucleosynthesis during a PIE, the age derived from cosmochronometry measures the time elapsed since the PIE in the AGB star.
This corresponds approximately to the time elapsed since the mass-transfer episode, or equivalently, since the onset of the white-dwarf cooling phase. Importantly, this is not the age of the observed CEMP-rs star itself, which is expected to be significantly older. This therefore provides a lower limit on the ages of both the CEMP-rs star and its white dwarf companion.
Such a nucleocosmochronometry is based on the comparison of production and measured abundance ratios of Th and/or U relative to one another or to a stable element (e.g. Eu). More specifically, the time spanned between production and present observation can be estimated from the expressions \citep{Cayrel2001}:
\begin{align}
\Delta t [\rm{Gy}]&= 46.7 \left[ \log (\mathrm{Th/Eu})_{\mathrm{ipro}} - \log \epsilon (\mathrm{Th/Eu})_{\mathrm{now}} \right] \\
\Delta t [\rm{Gy}]&= 14.8 \left[ \log (\mathrm{U/Eu})_{\mathrm{ipro}} - \log \epsilon (\mathrm{U/Eu})_{\mathrm{now}} \right] \\
\Delta t [\rm{Gy}]&= 21.8 \left[ \log (\mathrm{U/Th})_{\mathrm{ipro}} - \log \epsilon (\mathrm{U/Th})_{\mathrm{now}} \right]
\end{align}
where "ipro" refers to the theoretical i-process production ratio and "now" indicates the measured value. This age determination is obviously very sensitive to both abundance ratios. More specifically, an 0.1 dex uncertainty on the theoretical or measured abundance ratio of [Th/Eu] or [Th/U] inevitably induces an uncertainty on the $\Delta t$ age determination of 4.7 and 2.1~Gy, respectively. In principle, the Th/U ratio provides the most reliable radioactive chronometer since Th and U are neighbouring nuclei and their production factors are likely bound to the same stellar conditions \citep{Goriely16c}, hence reducing the theoretical uncertainties.

As discussed in \citet{Choplin2022letter,Martinet2024,Choplin2025}, the Th and U production by the i-process is still affected by many nuclear and astrophysics model and parameter uncertainties. In particular, when considering only the parameter uncertainties affecting the nuclear reaction rates derived by \citet{Martinet2024} and propagated into i-process simulation in AGB stars (see Sect.~\ref{Sect:nucleosynthesis}), we find possible variation of 2.5 and 1.1 dex in the [Th/Eu]$_{\rm ipro}$ and [Th/U]$_{\rm ipro}$, respectively. These translate into age uncertainties of more than 100~Gyr and 24~Gyr, respectively.
For this reason, there is no hope at the present time to extract a reliable age out of the Th and/or U chronometers before strongly reducing the nuclear uncertainties first. As shown by \citet{Martinet2024}, a significant contribution to these nuclear uncertainties stem from the still experimentally unknown $^{217}$Bi(n,$\gamma$)$^{218}$Bi branching rate.



\section{Conclusions}
\label{Sect: conclusion}



In this study, we report the detection of thorium in three CEMP-rs stars based on NLTE stellar parameters, confirming the presence of actinide elements in these objects. Uranium is not firmly detected, but upper limits are derived. We explored the production of these elements in the framework of i-process nucleosynthesis and find that the observed abundance patterns are more consistently reproduced by i-process AGB models than by pure s-process predictions. While alternative interpretations, such as the superposition of independent r- and s-process contributions, cannot be excluded, they are not investigated in this work.
Thorium and uranium can be produced by the i-process, although significant uncertainties in nuclear physics and stellar modelling still affect the predicted yields and production ratios. Consequently, age estimates based on Th/U nucleocosmochronometry remain highly uncertain, further limited by the absence of precise uranium abundances.
Overall, the detection of actinides in CEMP-rs stars provides new observational constraints on heavy-element nucleosynthesis and supports the relevance of the i-process in explaining their chemical peculiarities.

\begin{acknowledgements}

We thank the anonymous referee for the constructive suggestions. MR acknowledges the financial support from UGC, Govt. of India, through the UGC-JRF (NTA Ref.No:201610156431/CSIR-UGC NET NOVEMBER 2020). DK acknowledges Anusandhan National Research Foundation (ANRF), New Delhi for the award of a multi-institutional PAIR grant (ANRF/PAIR/2025/000021/ PAIR-A) to the University of Calicut. DK acknowledges the financial support from ANRF through the SURE grant with file number (SUR/2022/000748). MR and DK gratefully acknowledge financial support from the Belgium - India project on Precision Astronomical spectroscopy for Stellar and Solar system bodies" (BIPASS), approved by the International Division, Department of Science and Technology (DST, Govt. of India; DST/INT/BELG/P-01/2021 (G)) and the Belgian Federal Science Policy Office (BELSPO, Govt. of Belgium; BL/33/IN22\_BIPASS).  SVE thanks the ULB Foundation for their support.
SG and AC acknowledge financial support from F.R.S.-FNRS (Belgium). This work was supported by the F.R.S.-FNRS under Grant No. IISN 4.4502.19 and under the EOS Project No. O000422 and O022818F.

\end{acknowledgements}

\facilities{VLT:Kueyen (UVES)}


\appendix



\newpage
\begin{table}
\section{Line list}
The atomic lines used for deriving the elemental abundances are listed in Table~\ref{Tab:Linelists}. For all elements except Th and U, the departures from LTE are incorporated for the spectral lines while deriving the abundances.

\renewcommand{\thetable}{A\arabic{table}}
\setcounter{table}{0}
\caption{Atomic lines used for the abundance analysis.
\label{Tab:Linelists}}
\centering

\centering

\hfill
\begin{minipage}[t]{0.20\textwidth}
\vspace{10pt}

\begin{tabular}{l@{\hspace{8pt}}l@{\hspace{2pt}}r@{\hspace{2pt}}}

\toprule
$\lambda$ (\AA) & $\chi_{\rm low}$  & $\log gf$ \\
 & (eV) &      \\
\midrule \\
Fe I & & \\
3873.760 & 2.433 & -0.876\\
3893.390 & 2.949 & -0.602\\
4222.213 & 2.449 & -0.967\\
4442.832 & 2.176 &-2.792\\
4772.803 & 1.557 & -2.897\\
4772.830 & 3.017 & -2.192\\
4903.308 & 2.882 & -0.926\\
4985.253 & 3.929 &-0.447\\
4994.129 & 0.915 &-3.058\\
5005.712 & 3.884 & -0.120\\
5006.119 & 2.833 & -0.631\\
5074.748 & 4.220 & -0.230\\
5079.223 & 2.198 & -2.068\\
5166.282 & 0.000 &-4.192\\
5198.711 & 2.223  &-2.135\\
5202.336 & 2.176 &-1.838\\
5216.274 & 1.608 &-2.082\\
5217.389 & 3.211 & -1.074\\
5225.526 & 0.110 & -4.789\\
5242.491 & 3.634 & -0.967\\
5247.050 & 0.087 & -4.949\\
5250.209 & 0.121 & -4.933\\
5281.790 & 3.039 & -0.833\\
5288.525 & 3.695 & -1.493\\
5339.929 & 3.266 & -0.635\\
5364.871 & 4.446 & 0.228\\
5367.466 & 4.415 & 0.444\\
5383.369 & 4.313 & 0.645\\
5410.910 & 4.473 & 0.398\\
5415.199 & 4.387 & 0.643\\
5569.618 & 3.417 &-0.517\\
5572.842 & 3.397 &-0.289\\
5586.767 & 4.260 & -3.023\\
5753.122 & 4.260 & -0.623\\
5862.356 & 4.549 & -0.127\\
6003.011 & 3.882 &-1.100\\
6136.615 & 2.453  &-1.402\\
6137.691 & 2.588 &-1.402\\
6200.312 & 2.609 &-2.433\\
6219.280 & 2.198 &-2.432\\
6230.722 & 2.559 &-1.281\\
6252.555 & 2.404 &-1.699\\
\hline\\

\end{tabular}
\end{minipage}
\hfill
\begin{minipage}[t]{0.20\textwidth}
\vspace{10pt}
\begin{tabular}{l@{\hspace{8pt}}l@{\hspace{8pt}}r@{\hspace{2pt}}}
\toprule
$\lambda$ (\AA) & $\chi_{\rm low}$ & $\log gf$ \\
 & (eV) &      \\
\midrule\\
6335.330 & 2.198 &-2.177\\
6400.000 & 3.603 &-0.276\\
6411.648 & 3.654 & -0.596\\
6546.238 & 2.759 &-1.536\\
\hline\\
Fe II & & \\
3783.347 & 2.276 & -3.387\\ 
3938.290 & 1.671 & -4.073\\
4489.176 & 2.828 &-2.971\\
4491.405 & 2.856  &-2.756\\
4515.333 & 2.844 & -2.450\\
5197.567 & 3.230 & -2.220\\
5264.802 & 3.230 & -3.130\\
5284.103 & 2.891 & -3.195\\
5316.609 & 3.153 &-1.870\\
5316.781 & 3.221 & -2.740\\
5362.861 & 3.199 & -2.570\\
5534.838 & 3.245 & -2.865\\
6247.557 & 3.892 & -2.435\\
6432.676 & 2.891 &-3.570\\ 
\hline\\
O I && \\
6300.304 & 0.000 &-9.715\\
\hline\\
Na I && \\
5682.633 & 2.102 &-0.706\\
5688.205 & 2.104 &-0.404\\
6160.747 & 2.104  &-1.246\\
\hline\\
Mg I && \\
4702.991 & 4.346 & -0.456\\
5528.405 & 4.346& -0.547\\
5711.088 & 4.346 & -1.742\\
\hline\\
Ca I && \\
5581.965 & 2.523 &-0.555\\
5588.749 & 2.526 & 0.358\\
5590.114 & 2.521 & -0.571\\
5594.462 & 2.523 & 0.097\\
5598.480 & 2.521 &-0.087\\
6102.723 & 1.879 & -0.850\\
6122.217 & 1.886& -0.380\\
6162.173  &1.899& -0.170\\

\hline\\

\end{tabular}
\end{minipage}
\hfill
\begin{minipage}[t]{0.20\textwidth}
\vspace{10pt}
\begin{tabular}{l@{\hspace{8pt}}l@{\hspace{2pt}}r@{\hspace{2pt}}}
\toprule
$\lambda$ (\AA) & $\chi_{\rm low}$  & $\log gf$ \\
 & (eV) &      \\
\midrule\\

6169.563 & 2.526& -0.478\\
\hline\\
Ti II & & \\
4316.794 & 2.048 &-1.620\\
4320.950 & 1.165 &-1.880\\
4344.281 & 1.084 &-1.910\\
4417.713 & 1.165 & -1.190\\ 
4418.331 & 1.237 &-1.990\\
4443.801 & 1.080 & -0.710\\
4444.554 & 1.116 &-2.200\\
4450.482 & 1.084 &-1.520\\
4488.324 & 3.124  &-0.500\\
4501.270 & 1.116 & -0.770\\
\hline\\
Mn I &&\\
4030.753 & 0.000 & -0.470\\
4033.062 & 0.000  &-0.618\\
4754.047 & 2.282 & -0.668\\
4783.405 & 2.298 & -1.375\\
\hline\\
Co I &&\\
4118.767 & 1.049 & -0.490\\
4121.311 & 0.923 & -0.320\\
\hline\\
Ni I &&\\
5035.357 & 3.635 & 0.290\\
5081.110 & 3.847 & 0.462\\
5146.482 & 3.706 & 0.060\\
5476.903 & 1.826 & -0.780\\
\hline\\
Sr I &&\\
4607.331 & 0.000 & 0.283\\
\\
Sr II &&\\
4077.719 & 0.000 & 0.170\\
4215.519 & 0.000 &-0.170\\
\hline\\
Y II &&\\
4823.304 & 0.992 &-0.976\\
4854.861 & 0.992& -0.270\\
4883.682 & 1.084 & 0.190\\
4900.119 & 1.033 & 0.030\\
4982.129 & 1.033 &-1.320\\

\hline\\
\end{tabular}
\end{minipage}
\hfill
\begin{minipage}[t]{0.20\textwidth}
\vspace{10pt}
\begin{tabular}{l@{\hspace{8pt}}l@{\hspace{8pt}}r@{\hspace{2pt}}}
\toprule
$\lambda$ (\AA) & $\chi_{\rm low}$  & $\log gf$ \\
 & (eV) &      \\
\midrule \\

5087.416 & 1.084 &-0.160\\
5200.406 & 0.992 &-0.470\\
5205.722 & 1.033& -0.280\\
5402.774 & 1.839& -0.310\\
5509.895 & 0.992& -1.310\\
\hline\\
Ba II &&\\
4166.000 & 2.722 &-0.420\\
4524.925 & 2.512 &-0.360\\
5853.676 & 0.604 &-1.965\\
\hline\\
Eu II &&\\ 
3819.661 & 0.000 & 0.190\\
3907.117 & 0.207 &-0.112\\
3930.509 & 0.207& -0.012\\
3971.983 & 0.207 &-0.012\\
4129.720 & 0.000 &-0.062\\
6437.679 & 1.320 &-1.037\\
6645.098 & 1.380 &-0.319\\
\hline\\
Th II  &&\\
3433.999 & 0.231 & -0.537\\
3435.977 & 0.000  &-0.670\\
3469.921 & 0.514 & -0.129 \\
3539.587 & 0.000 & -0.542\\
4019.129 & 0.000 & -0.228 \\
4086.521 & 0.000 &-0.929\\
\hline\\
U  II  &&\\
3859.571 & 0.036 & -0.067\\
4050.041 & 0.000 & -0.706 \\
4090.132 & 0.217  & -0.184 \\
\hline\\
\end{tabular}
\end{minipage}
\hfill

\end{table}


{
\begin{table*}[h!]
\renewcommand{\thetable}{B\arabic{table}}
\setcounter{table}{0}
\section{Elemental abundances for the program stars}
\caption{Elemental abundances
\label{Tab:abundances}}
\hspace*{-3cm}
\begin{tabular}{lclrrrrrrrrrrrrrr}
\hline
\multicolumn{2}{c}{}& \multicolumn{5}{c}{HD~187861} && \multicolumn{4}{c}{HD~224959} && \multicolumn{4}{c}{CS~22891-171} \\
\cline{4-7}\cline{9-12} \cline{14-17}\\
 &    Z  &    log$_{\odot}{\epsilon}^a$ & log${\epsilon}$&$\sigma_{l}$ (N)& [X/Fe] $\pm~\sigma_{t}$ & & & log${\epsilon}$&$\sigma_{l}$(N)& [X/Fe]~$\pm~ \sigma_{t}$ & & &  log${\epsilon}$&$\sigma_{l}$(N)& [X/Fe]~$\pm~ \sigma_{t}$ &  \\
\hline\\
C  & 6  & 8.43 &   8.40  &   0.10(2)  &   2.37 $\pm$ 0.17  && & 7.90 &  0.04(3) &  1.73 $\pm$ 0.16   & & & 7.75  &   0.12(3)  &   1.51 $\pm$ 0.17  & \\
$^{12}$C/$^{13}$C$^c$&-- &-- & -- &    --   &  15.6 $\pm$ 3.3   &  & &      --   &     --      &  11.5 $\pm$      1.9  & & & -- &  --  &  2.33 $\pm$ 0.7   & \\
N  & 7  & 7.83 &   8.00  & 	 0.10(1)  &   2.57 $\pm$ 0.15 && &    7.50  & 	 0.09(29) &   1.93 $\pm$ 0.12  && &  7.20  & 	 0.10(1)  &   1.56 $\pm$ 0.15  & \\
O$_{\rm NLTE}$  & 8  & 8.69 &    7.80    &   0.10(1)    &    1.51 $\pm$ 0.17 && &     7.50  &   0.10(1)  &  1.07 $\pm$ 0.16  &&&--&--&--\\
Na$_{\rm NLTE}$ & 11 & 6.24 &   4.15  &   0.10(1)  &  0.31 $\pm$ 0.12  & &&   4.15&  0.10(2)  &  0.17 $\pm$ 0.11   &  & & 5.10 &   0.10(2)  &  1.05 $\pm$ 0.13   &\\
Mg$_{\rm NLTE}$ & 12 & 7.66 &   5.65  &   0.10(1)  &  0.39 $\pm$ 0.13   & &&    5.60 &   0.10(1) & 0.20 $\pm$ 0.14  & &&   5.75  &   0.05(2)  &  0.34 $\pm$ 0.13   &   \\

Ca$_{\rm NLTE}$ & 20 & 6.34 &   4.39  &   0.17(4)  &  0.45 $\pm$ 0.11   & &&     4.38  &   0.12(6)  & 0.30 $\pm$ 0.12   & & &  4.50  &   0.16(7)  &  0.35 $\pm$ 0.11   & \\
Sc & 21 & 3.15 &   1.45  &   0.10(1)  &  0.70 $\pm$ 0.13   & &&  1.18 &   0.11(3)  &  0.29 $\pm$ 0.12   & && 1.20 &  0.10(1)   & 0.24 $\pm$ 0.13  &    \\
Ti$_{\rm NLTE}$ & 22 & 4.95 &   2.60  &   0.10(2)  &  0.05 $\pm$0.10   & &&     2.95  &   0.16(8)  &  0.26 $\pm$ 0.14  & &&  2.59  &   0.18(4)  &  $-$0.17 $\pm$ 0.17  &  \\
Cr & 24 & 5.64 &   3.05  &   0.04(3)  &  $-$0.19 $\pm$ 0.16   & & &   3.26 &   0.09(8)  & $-$0.12 $\pm$ 0.14  & &&  3.19 &   0.21(4)  &  $-$0.26 $\pm$ 0.15   &   \\
Mn$_{\rm NLTE}$ & 25 & 5.43 &   3.00  &   0.10(1)  &  $-$0.03 $\pm$ 0.12  && &     3.20 &   0.10(2) & 0.03 $\pm$ 0.11    & &&  2.28  &   0.02(2)  &  $-$0.96 $\pm$ 0.11 &   \\
Fe I& 26 & 7.50 &   5.11  &   0.15(25) &  -- & & &      5.27 &  0.11(40)  & -- &  & &   5.33  & 0.12(32)  & --&\\
Fe II& 26 & 7.50 &   5.04  &   0.15(9) & --  & & &      5.21 &  0.14(11)  & -- &  & &   5.22  & 0.13(8)  & --&\\
Fe & 26 & 7.50 &   5.10  &   0.15(34) & --  & & &      5.24 &  0.12(51)  & -- &  & &   5.31  & 0.14(40)  & --&\\
Co$_{\rm NLTE}$ & 27 & 4.99 &   2.80 &   0.10(1)  &  0.21 $\pm$ 0.12   &&  &     3.00&   0.10(1) & 0.27 $\pm$ 0.12   & &&  2.40 &   0.05(2)  &  $-$0.40 $\pm$ 0.16 &  \\
Ni$_{\rm NLTE}$ & 28 & 6.22 &   4.40  &   0.10(1)  &  0.58 $\pm$ 0.11   & &&  3.95 &  0.05(2) & $-$0.01 $\pm$ 0.12   & && 3.82 &  0.02(2)  & $-$0.21 $\pm$ 0.12 &  \\
Cu & 29 & 4.19 &  2.70  & 0.10(1) & 0.91 $\pm$ 0.14  & & &   1.80 &  0.10(1)   &  $-$0.13 $\pm$ 0.14  & && --&--& --& \\
Zn & 30 & 4.56 &   2.30 &   0.10(1)  &  0.14 $\pm$ 0.14  && &    2.50&   0.10(2) &  0.20$\pm$ 0.15    & &&  2.47 &   0.02(2)  &  0.10 $\pm$ 0.12 &   \\
Sr$_{\rm NLTE}$ & 38 & 2.87 &  0.60  &   0.10(1)     &   0.13 $\pm$ 0.13      && &  1.40  &  0.10(1)    &   0.79 $\pm$ 0.13     & &&   1.40:    &   0.10(2)      &  0.72: $\pm$ 0.15 &  \\
Y$_{\rm NLTE}$  & 39 & 2.21 &   0.45  &   0.04(4)  &  0.64 $\pm$ 0.09  & &&   0.52  &   0.07(6)  &  0.57 $\pm$ 0.08  & && 0.92  &   0.16(8)  &  0.90 $\pm$ 0.12 &\\
Zr & 40 & 2.58 &   1.20  &   0.10(1)  &  1.02 $\pm$ 0.11  && &   1.50 &  0.11(4)  &  1.19 $\pm$ 0.15   &&&  1.75 &  0.32(4)  &  1.36 $\pm$ 0.13 & \\
Nb & 41 & 1.46 &  0.60  &     0.10(2)     &  1.60 $\pm$ 0.12  && &     0.60 &       0.10(1) &   1.40 $\pm$ 0.12    & &&  0.75 &      0.05(2)     &  1.48 $\pm$ 0.17   &  \\
Ba$_{\rm NLTE}$ & 56 & 2.18 &   1.90  &   0.10(1)  &  2.12 $\pm$ 0.14  & &&   1.62 &  0.10(1)  &  1.70 $\pm$ 0.14  &&&  1.60  &   0.10(1)  &  1.61 $\pm$ 0.14 & \\
La & 57 & 1.10 &  0.65  &   0.06(6) &  1.95 $\pm$ 0.09   & &&     0.90 & 0.08(10) & 2.06 $\pm$ 0.08  && & 1.04  & 0.05(16) & 2.13 $\pm$ 0.11 & \\
Ce & 58 & 1.58 &  1.00 &   0.08(3) &  1.82 $\pm$ 0.13   & &&    1.25 &  0.09(10) &  1.92 $\pm$ 0.13  &&& 1.26 &   0.10(5) &  1.87 $\pm$ 0.15 &  \\
Pr & 59 & 0.72 &  0.30  &   0.12(5)  & 1.98 $\pm$ 0.14  & &&     0.56&   0.11(5)  & 2.10 $\pm$ 0.13  &&&  0.54 &   0.06(5)  &2.01 $\pm$ 0.12 & \\
Nd & 60 & 1.42 &   0.95  &   0.07(11) &  1.93 $\pm$ 0.11  && &   1.33  &  0.07(15) & 2.17 $\pm$ 0.11  & &&  1.24  & 0.07(14) & 2.01 $\pm$ 0.14 &  \\
Sm & 62 & 0.96 &  0.15  &   0.05(2)  &  1.59 $\pm$ 0.14   & &&  0.70 &   0.11(8) &  2.00 $\pm$ 0.12  &&&  0.60&  0.10(2)   &  1.83 $\pm$ 0.13 &    \\
Eu$_{\rm NLTE}$ & 63 & 0.52 &  $-$0.15  &   0.07(3)  & 1.80 $\pm$ 0.16   &&& $-$0.17 & 0.05(3)   & 1.57 $\pm$ 0.14    &  & &   $-$0.03  &  0.14(4)  & 1.64 $\pm$ 0.15  & \\
Gd & 64 & 1.07 &   0.10 &   0.10(1)  & 1.43 $\pm$ 0.12   & &&  0.30 &   0.10(1)  & 1.49 $\pm$ 0.12   & & &  0.52 &   0.07(2)  & 1.64 $\pm$ 0.13 & \\
   
Tb & 65 & 0.30 &$-$0.90  &   0.10(1)  &  1.20 $\pm$ 0.12 && &  $-$0.70  & 0.08(3) & 1.26 $\pm$ 0.14  &  &&  $-$0.37  &  0.02(2)  & 1.52 $\pm$ 0.13 & \\
Dy & 66 & 1.10 &   $-$0.10 &   0.10(2)  & 1.20 $\pm$ 0.11   &&&  0.15 &   0.05(2)  & 1.31 $\pm$ 0.11  &&&  0.00 &   0.02(3)  & 1.09 $\pm$ 0.13 &\\
Ho & 67 & 0.48 &   $-$0.77 &   0.09(3)  & 1.15 $\pm$ 0.17   & &&  $-$0.40: & 0.08(3)  & 1.38: $\pm$ 0.16  &&&   $-$0.69 &  0.11(2)  & 1.02 $\pm$ 0.15 & \\
Er & 68 & 0.92 &  $-$0.30 &   0.07(4)  &  1.18 $\pm$ 0.11    && &   0.77&  0.12(3) &  2.11$\pm$ 0.14 & && $-$0.03 & 0.11 (4)  &  1.24 $\pm$ 0.13 &  \\
Tm & 69 & 0.10 &   $-$0.95 &   0.05(2)  & 1.35 $\pm$ 0.12  & &&    $-$0.68 &  0.07(3)  & 1.48 $\pm$ 0.12  &&&   $-$0.73 &  0.04(3)  & 1.36 $\pm$ 0.12 &\\
Yb & 70 & 0.84 &   $-$0.40: &   0.10(1)  & 1.16: $\pm$ 0.15  & &&   0.40: &   0.10(1)  & 1.82: $\pm$ 0.15 &&&   $-$0.40: &   0.10(1)  & 0.95: $\pm$ 0.15   & \\
Lu & 71 & 0.10 &   $-$0.60: &   0.10(1)  & 1.70: $\pm$ 0.17  & &&   $-$0.15: &   0.05(2)  & 2.01: $\pm$0.16 &&&--&--& --&\\
Hf & 72 & 0.85 &   0.30  &   0.10(2)  & 1.85$\pm$ 0.12  & &&  0.56 &  0.12(3)  & 1.97 $\pm$ 0.15  & &&  0.38 &  0.02(2)  & 1.72 $\pm$ 0.11 &\\
Ta & 73 & $-$0.12 &  $-$0.30*  &  0.10(2)  & 2.22*$\pm$ 0.12  & &&   $-$0.20*  &  0.10(1)   &  2.18* $\pm$ 0.12 & &&  $-$0.20:  &  0.10(1)  & 2.11: $\pm$ 0.12 & \\
Os & 76 & 1.40 &  -- &  --  & --  & &&   0.30  &  0.10(1)   &  1.16 $\pm$ 0.17 & &&   0.50    &  0.10(1)   &    1.29 $\pm$ 0.17   &\\
Ir & 77 & 1.38 &     1.20:   &  0.10(1)  & 2.22: $\pm$ 0.18 &     & &   1.50:: & 0.10(1)& 2.38:: $\pm$ 0.18 &&&  0.75:   &  0.15(2)  & 1.56:$\pm$ 0.19  &\\
Pb & 82 & 1.75 &   2.50 &   0.10(1)  &  3.15 $\pm$ 0.12  && &  2.40&  0.10(1)  &  2.91$\pm$ 0.12    &  && 1.65&  0.10(1)  &  2.09$\pm$ 0.12 & \\
Th & 90 & 0.02 &   $-$0.76 &   0.04(3)  &  1.62 $\pm$ 0.14  && &  $-$0.90&  0.05(2)  &  1.34 $\pm$ 0.15    &  &&  $-$0.63 &   0.06(4)  &  1.54 $\pm$ 0.17  &\\
U & 92 & $-$0.54 &    $-$1.40* &   0.10(1)  &  1.54* $\pm$0.18  && &  $-$1.45*&  0.10(1)  &  1.25*$\pm$ 0.18    &  &&   $-$1.33* &   0.02(3)  &  1.40* $\pm$ 0.14  &\\
\hline\\
\end{tabular}

$^{a}$ Asplund et al. (2009) \\
$:$ Uncertain abundances due to noisy/blended region\\
$::$ Very uncertain abundances due to noisy/blended region\\
$*$ Upper limit and uncertain\\
$^{c}$ $^{12}$C/$^{13}$C ratio from CH G band\\
\end{table*}
}


\begin{table}
\section{Uncertainties in Elemental Abundances}
The uncertainties associated with the elemental abundances derived for the program stars are presented in Table~\ref{Tab:abundances}, were estimated following the procedure outlined in Section 4.6 of \citet{Riyas2026}. The quantities $\sigma_{T}$, $\sigma_{\log g}$, and $\sigma_{\xi}$  are the typical uncertainties on the atmospheric parameters, which are estimated to be  $\sigma_{T}$ = 50~K, $\sigma_{\log g}$ = 0.2~dex, $\sigma_{\xi}$ = 0.05~km/s, and $\sigma_{\mathrm{[Fe/H]}}$ = 0.10~dex. The partial derivatives were determined using HD~187861 for all elements except Os, for which HD 224959 was adopted, varying the atmospheric parameters $T_{\rm eff}$, $\log g$, microturbulence $\xi$, and [Fe/H] by 100~K, 0.5, 0.5~km/s, and 0.5 dex, respectively and the corresponding abundance changes are provided in Table ~\ref{Tab:uncertainties}. The covariance terms were directly measured for HD 187861, yielding   $\sigma_{T \log g}$ = 12, $\sigma_{\log g \xi}$ = $-$0.01 and $\sigma_{\xi T}$ = 6.

\centering
\renewcommand{\thetable}{C\arabic{table}}
\setcounter{table}{0}
\caption{ Abundance variations ($\Delta \log \epsilon_{X}$) with variations of the atmospheric parameters.}
\label{Tab:uncertainties}
\begin{tabular}{crrrr}
\hline
       & \multicolumn{4}{c}{$\Delta \log \epsilon_{X}$} \\
        \cline{2-5}\\
 Element&   $\Delta T_{\rm eff}$ & $\Delta \log g$ & $\Delta \xi_t$ &$\Delta$ [Fe/H]   \\
&   ($+$100 K) & ($+$0.5) & ($+$0.5  & ($+$0.5 \\
   &        &         & km~s$^{-1}$)& dex) \\
\hline\\
C  &  0.15  &  0.08   &    $-$0.05  &   $-$0.01 \\
N  &  0.18  &  0.04   &     0.02  &    0.02 \\
O  &  0.03  &  $-$0.20   &    0.05  &   0.00 \\
Na &  0.02  &  $-$0.15   &     0.01  &    $-$0.09 \\
Mg &  0.03  &  $-$0.80   &     0.00  &    0.00 \\
Ca &  0.12  &  $-$0.07  &    0.01  &    $-$0.02 \\
Sc &  0.04  &  0.18   &     0.00  &    $-$0.01 \\
Ti &  0.08  &  $-$0.06   &     $-$0.01  &    0.00 \\
Cr &  0.12  &  $-$0.02   &     $-$0.09  &    $-$0.02 \\
Mn &  0.10  &  $-$0.18   &     0.05  &    0.00 \\
Fe &  0.05  &  $-$0.07   &    $-$0.11  &    $-$0.21 \\
Co &  0.13  &  $-$0.06   &     $-$0.05  &    $-$0.06 \\
Ni &  0.14  &  $-$0.03   &     $-$0.04  &    0.02 \\
Cu &  0.07  &  $-$0.10   &     0.03  &    0.01 \\
Zn &  0.07  &  $-$0.20   &     0.06  &    0.03 \\
Sr &  0.10  & $-$0.15   &    $-$0.04  &    $-$0.02  \\
Y  &  0.07  &  0.16   &     $-$0.01  &   $-$0.03 \\
Zr &  0.08  &  0.14   &    $-$0.04  &   $-$0.03 \\
Nb & 0.05  &  0.01   &$-$0.03   &  $-$0.02 \\
Ba &  0.09  &  0.20   &     $-$0.17  &    $-$0.18 \\
La &  0.06  &  0.12   &    0.00  &   $-$0.01 \\
Ce &  0.06  &  $-$0.01   &   0.02  &   0.00\\
Pr &  0.06  &  0.09   &    0.00  &   $-$0.02 \\
Nd &  0.07  &  0.02   &    $-$0.01  &   $-$0.02  \\
Sm & 0.08  &  0.03   &   0.02  &   0.00 \\
Eu & 0.06  &  $-$0.02   &  0.02  &   $-$0.01 \\
Gd &  0.10  &  0.17   &   0.00  &   $-$0.02\\
Tb &  0.15  &  $-$0.60   &    0.06  &   0.00 \\
Dy &  0.07  &  $-$0.11   &     $-$0.01  &    0.00 \\
Ho &  0.13  &  0.12   &     $-$0.01  &    0.00 \\
Er &  0.12  &  0.11   &    $-$0.09  &   $-$0.01 \\
Tm &  0.07  &  0.04   &     0.01  &    $-$0.02 \\
Yb &  0.14  &  0.22   &     $-$0.04  &    $-$0.01 \\
Lu &  0.09  & $-$0.60    &  $-$0.02  & $-$0.01 \\
Hf &  0.08  &  0.04   &   0.00  &   0.00 \\
Ta &  0.25 &  $-$0.04   &    0.06  &   0.08 \\
Os &  0.37  &  $-$0.09   &    0.01  &   $-$0.07 \\
Ir &  $-$0.05  & 0.04   & 0.03  &0.01 \\
Pb &  0.19  &  0.00   &     $-$0.08  &    $-$0.04 \\
Th &  0.30  &  $-$0.60   &     0.12  &    0.05 \\
U &  0.25  &  0.50   &     0.11  &    0.00 \\
\hline
\end{tabular}
\end{table}

\clearpage


\bibliography{CEMP-ref}
\bibliographystyle{aasjournalv7}



\end{document}